\documentclass[a4paper,10pt,twoside]{cpc-hepnp}

\usepackage{multicol}
\usepackage{graphicx}
\usepackage{booktabs}
\usepackage{amssymb,bm,mathrsfs,bbm,amscd}
\usepackage[tbtags]{amsmath}
\usepackage{lastpage}
\usepackage{CJK}
\usepackage{epstopdf}
\usepackage{xcolor}
\begin{document}
\begin{CJK*}{GBK}{song}

\fancyhead[c]{\small }
\fancyfoot[C]{\small 010201-\thepage}

\footnotetext[0]{Received }

\title{Lorentz Violation, Quantum Tunneling and Information Conservation}
\author{%
Guo-Ping Li$^{1;1)}$\email{Corresponding author, gpliphys@yeah.net}, Ke-Jian He$^{1;2)}$\email{kjhe94@163.com}, Bing-Bing Chen $^{2;3)}$\email{binbchen@126.com} %
}
\maketitle
%\author{%

\address{%
$^1$ Physics and Space College, China West Normal University, Nanchong 637000, People's Republic of China
}

\address{%
$^2$ Department of Physics, Si Chuan MinZu College, Kangding, 626001, People's Republic of China}

\begin{abstract}
  In this paper, by introducing a Lorentz-invariance-violation (LIV) class of dispersion relations (DR) suppressed by the second power $(E/E_{QG})^2$, we have investigated the effect of LIV on the Hawking radiation of the charged Dirac particle via tunneling from a Reissner-Nordstr\"{o}m(RN) black hole. We first find the effect of LIV speeds up the black hole evaporation, leaving the induced Hawking temperature very \emph{sensitive} to the changes in the energy of the radiation particle, but at the same energy level, \emph{insensitive} to the changes in the charge of the radiation particle. This provides a phenomenological evidence for the LIV-DR as a candidate for describing the effect of quantum gravity. Then, when the effect of LIV is included, we find the statistical correlations with the Planck-scale corrections between the successive emissions can leak out the information through the radiation. And, it turns out that the black hole radiation as tunneling is an entropy conservation process, and no information loss occurs during the radiation, where the interpretation for the entropy of black hole is addressed. Finally, we conclude that black hole evaporation is still an unitary process in the context of quantum gravity.
\end{abstract}

\footnotetext[0]{\hspace*{-3mm}\raisebox{0.3ex}{$\scriptstyle\copyright$}2020
Chinese Physical Society and the Institute of High Energy Physics
of the Chinese Academy of Sciences and the Institute
of Modern Physics of the Chinese Academy of Sciences and IOP Publishing Ltd}%

\begin{multicols}{2}
\section{Introduction}
\label{sec:intro}
\setlength{\parindent}{2em}
Since the observable signals of Lorentz invariance violation(LIV) can be described by using effective field theory\cite{LIV+1}, the Planck-scale physics effect induced by the LIV has been extensively studied in the past decades\cite{Giovanni1,Giovanni5,Giovanni0,Townsend}. In the standard model extension (SME), Colladay and Alan Kostelec\'{k} have observed that the spontaneous LIV would occur in the low-energy limit of a physically relevant fundamental theory\cite{LIV+2}. And in 2002, the fact that the Double Special Relativity (DSR) naturally leads to the Lorentz invariance violation of dispersion relations (LIV-DR) has been presented by Amelino-Camelia\cite{Giovanni2}. More importantly, it is generally believed that the introduction of gravity into quantum theory would give rise to the Planck-scale departure from the Lorentz symmetry\cite{LIV+3,LIV+4,LIV+5,Giovanni3,Giovanni6,Alfaro}. In string theory, the potential mechanism for spontaneous breakdown of Lorentz symmetry has been detailed via covariant string field theory\cite{LIV+3}. The possible violations of Lorentz symmetry motivated by loop quantum gravity have been carefully discussed by considering the correspondence principle\cite{LIV+4}. And in Ho\v{r}ava's-Lifshitz theory, it is true that the violation of Lorentz invariance is also included\cite{LIV+5}. In view of this, more and more experimental observations have implied that the LIV might be regarded as an effective model for exploring the effect of quantum gravity\cite{GRB1,GRB2,GRB3,GRB7,Jacobson,Jacob}.
With the aid of the LIV, various constraints on the assumed energy scale for quantum gravity effects $E_{QG}$ have been reported from the Gamma-ray bursts (GRBs)\cite{GRB1,GRB2,GRB3,GRB7}. In the context of the LIV-DR, the observation of $100-MeV$ synchrotron radiation from the Crab nebula provides an important constraint on theories of quantum gravity\cite{Jacob}. Moreover, there are many other similar and remarkable phenomenons, which are designed to investigate the violation of Lorentz invariance in experiment\cite{Other1,Other2,Other3,Other4,Other5,Other6,Other7,Other8,Other9,Other10,Giovanni4,LIVEX,LIVEX1}, i.e.,  testing Lorentz-symmetry violation with Atomic Systems\cite{LIVEX,LIVEX1}. In a word, all of these observations show that the LIV as a candidate for describing the effect of quantum gravity has been one of the most interesting and hottest topics in recent years.

From the phenomenological perspective, we in this paper adopt the simple framework of deformed dispersion relation to characterize the Lorentz violation. Assuming the preferred frames in which dispersion relation break boost invariance but preserve rotation invariance, a generic approximate quantum-gravity-induced LIV-DR at high-energy regime can be expected to the following form (\ref{Le2})\footnote{For convenience, we have chosen the form (\ref{Le2}) to work in our paper, other different forms of LIV-DR can also be found in the Standard Model Extension (SME)\cite{add1}.} \cite{GRB1,GRB2,GRB3,GRB7,Jacobson,Jacob,Giovanni3,Giovanni6},

\begin{equation}\label{Le2}
{E^2} = {p^2} + {m^2} - {\eta_ \pm }{p^2}{\left( {\frac{E}{{{\xi_n}{M_{QG}}}}} \right)^n},
\end{equation}
where, we only consider the leading quantum-gravity correction of the LIV in Eq. (\ref{Le2}). It should be noted that this relation can be considered only when it occurs at high energy scales, where $E$, $m$ and $p$ are the energy, mass and the momentum of the particles, respectively. As described in our previous work\cite{LGP1}, we have enough evidence to find that the parameters $E_{QG}$, $\eta_ \pm$ and $n$ should be properly fixed to $E_{p}$, $1$ and $2$, where $E_{p}$ is the Planck energy scale. Therefore, the LIV-DR can be usually expressed as the following form\cite{Jacob,Jacobson}, it reads,
\begin{equation}\label{DDR}
{E^2} = {p^2} + {m^2} - {l_p}{p^2}{E^2},
\end{equation}
where ${l_p} = 1/({\xi_2}^2 {M_p}^2) =\frac{{L_p}^2}{{\xi_2}^2}$ is related to the Plank length, and $\xi_2$ is a dimensionless parameter.
\\
\indent

It is well-known that black hole as a very amazing object in our universe is always regarded as a test bed for a full theory of quantum gravity. And at the final stage of the black hole evaporation, the effect of quantum gravity induced LIV is so large that it must be taken into account during the black hole evolution. Considering those facts, one can see that it is very necessary for us to study the LIV effects on black hole thermodynamics.
In addition, gravity as a special interaction force, is very sensitive to the mass of the particle, i.e. the energy of the particle according to the mass-energy relation $\omega=mc^2$. This phenomenon is distinguished from several other interaction forces (i.e. electromagnetic interaction, strong interaction and weak interaction). In this paper, by studying the sensitivity of the LIV-DR induced Hawking temperature to the energy and charge of the emission particle, we attempt to find a phenomenological evidence for the LIV-DR as a candidate for describing the effect of quantum gravity.
On the other hand, quantum tunneling is very successful as a model for describing the black hole radiation. Basing on it, one can find the exact emission spectrum of the black hole deviates from the pure thermal spectrum, but is consistent with an underlying unitary theory. This provides a qualitative explanation for the puzzle of the black hole information loss \cite{Chen1,Jiang11,Angheben,HawkingNa,KrausNPB}. In recent work, it has been further shown that the Hawking radiation as tunneling is indeed an entropy conservation process, and no information loss occurs during the radiation \cite{ZhangPLB,ZhangPLB1,ZhangPLB2}\footnote{Of course, there are many other attempts that have been made to study this issue in the past decade \cite{infor1,infor2,infor3} and references therein.}. However, as far as I know, most of these interesting work on the Hawking radiation as tunneling has been focused on the semiclassical case, ignoring the effect of quantum gravity.
In fact, as stated above, this ignorance can not really describe the radiation process of the black hole, especially can not accurately describe the final stage of the black hole evaporation where the energy of the emission particle is very high. So, the effect of quantum gravity should be included throughout the correct description of the black hole radiation \footnote{At present, there is as yet no complete quantum theory of gravitation, so it is generally believed that the generalized uncertainty principle (GUP)\cite{LHL1,LHL2,LHL3} and the LIV-DR can be regarded as the effective model in the study of the effect of quantum gravity \cite{Giovanni3,Giovanni6,GRB2,GRB3,QG1,QG2,ChenT,QG3,QG6,QG7,QG8,QG9,Other1,Other2,Other3,Other4,Other5,Other6,Other7,Other8,Other9,Other10,Giovanni4,
Battisti,Lidsey,Hammad,Gangopadhyay,Ghosh,Myung1,Balasubramanian,Pedram,Pikovski,Das}.}.
Now, it is necessary to check the previous work about the Hawking radiation as tunneling within the inclusion of the effect of quantum gravity.
Specifically, by using the LIV induced Dirac equation, we will study the quantum gravity effect on the Hawking radiation of the charged fermions via tunneling from the R-N black hole. In addition, the modified dispersion relation(MDR) nears the minimum measurable length can also be treated as a quantum gravity candidate since the minimum length is a common feature of quantum gravity theories. In recent years, many works are devoted to study the effect of MDR\cite{KamaliGRG,MajhiPLB,MDR+1,MDR+2,MDR+3,MDR+4}. Considering the MDR, the generation of primordial perturbation in various cosmological evolutions was carefully addressed\cite{MDR+1}. Based on the Friedmann-Robertson-Walker(FRW) universe, the form of the MDR in theories with extra dimensions was also obtained\cite{MDR+2}. In particular, we note in \cite{KamaliGRG} that a new form of MDR with inclusion of a minimum length and a maximum momentum has been introduced to investigate the radiation of the R-N black hole. Obviously, in the LIV and MDR, effects of quantum gravity are shown from different perspectives, so it is interesting for us to compare with their results.
Furthermore, in the context of the generalized uncertainty principle (GUP)(i.e., another candidate for quantum gravity), Nozari and Saghafi have studied the puzzle of the information loss during the emission process, and obtained the non-zero correlations with the Plank-scale corrections between the successive emissions, but they are not adequate by themselves to recover information \footnote{In Nozari's analysis, it should be pointed out that the influence of conditional probability and the question that whether the entropy and information are conserved or not were lacking. And in previous studies \cite{lf1}, when one ignored the influence of conditional probability, their results are presented to be misleading in calculation of the statistical correlation.} \cite{NozariJHEP}. In this paper, by analysing a dynamic evolution behavior of the Dirac particle within the inclusion of the effect of quantum gravity (i.e. the effect of the LIV), we also attempt to check whether the statistical correlations with the Planck-scale corrections between the successive emissions can leak out the information through the radiation, and the black hole radiation as tunneling is an entropy conservation process, and no information loss occurs during the radiation.
\\
\indent
The remainders of the present paper are outlined as follows. In Sec.2, by considering the effect of the LIV-DR, we rewrite the dynamic Dirac equation to obtain the Hawking radiation of the charged fermions via tunneling from the R-N black hole, and analyse the sensitivity of the LIV-DR induced Hawking temperature to the energy and charge of the emission particle. Basing on the LIV-DR induced tunneling radiation, Sec.3 is devoted to obtain the statistical correlations with the Planck-scale corrections between the successive emissions, and check the black hole radiation as tunneling is an entropy conservation process, and no information loss occurs during the radiation. Sec.4 ends up with a brief discussion and conclusion.

\section{Charged fermions' tunneling from the R-N black hole}\label{sec2}

In this section,  we will study the effect of the LIV on the Hawking radiation of the charged Dirac particle via tunneling from the R-N black hole. At first,
it is necessary to obtain the dynamic Dirac equation by considering the effect of the LIV. According to the spirit of the LIV-DR,
the corrected Dirac equation can be rewritten as\cite{KruglovPLB},
\begin{equation}\label{q1}
\left[{{{\overline \gamma }^\mu}{\partial_\mu} + \overline m - il_p^{1/2}\left( {{{\overline \gamma }^t}{\partial_t}} \right)\left({{{\overline \gamma  }^j}{\partial_j}} \right)}\right]\Psi = 0.
\end{equation}
Obviously, the Lorentz symmetry is broken by the additional term ($ l_p $) under the boost transformation. In \cite{KruglovPLB}, Eq.(\ref{q1}) has been proved to be compatible with the quadratically-suppressed LIV-DR (\ref{DDR}) when one substituted the wave function $ \Psi \left( x \right) = \Psi \left( p \right)\exp [i\left( {\overrightarrow p \overrightarrow x - {p_0}{x_0}} \right)] $ into the corrected Dirac equation (\ref{q1}). And, $ {\overline \gamma^\mu} $ is the ordinary gamma matrix, $ \mu, j $ are the spacetime coordinates and space coordinates, respectively. It is obvious that, the corrected Dirac equation in curved spacetime should be of the form
\begin{equation}\label{q2}
\left[{{\gamma ^\mu }{D_\mu } + \frac{m}{\hbar } - i\hbar l_p^{1/2}\left( {{\gamma ^t}{D_t}} \right)\left( {{\gamma ^j}{D_j}} \right)} \right]\Psi  = 0,
\end{equation}
where $ m $ is the mass of the emission particle. In the curved spacetime, $ {D_\mu } $ represents $ {D_\mu } = {\partial_\mu } + {\Omega_\mu } + (i/\hbar) e A_\mu$, and $\gamma^\mu $ is the gamma matrix that satisfies the relation $ \left\{ {{\gamma^\mu },{\gamma^\nu }} \right\} = {\gamma^\mu }{\gamma^\nu } + {\gamma^\nu }{\gamma^\mu } = 2{g^{\mu \nu }}I $. $ e{A_\mu } $ and $ {\Omega _\mu } $ are the charge term and spin connection respectively. Next, considering the effect of the LIV-DR, we attempt to investigate the charged fermions' tunneling from the R-N black hole. For the R-N black hole, it is written as $
ds^2 = -f(r)dt^2 +g(r)^{-1}dr^2 +r^2 (d\theta^2 +\sin{\theta}^2 d\varphi^2)
$ with $
f(r) = g(r) = 1- \frac{2M}{r} +\frac{Q^2}{r^2} = \frac{(r -r_+)(r -r_-)}{r^2}
$,
where $ A_\mu = (A_t,0,0,0) = (\frac{Q}{r},0,0,0) $ represents the electromagnetic potential, $ r_{\pm} =M \pm \sqrt{M^2 - Q^2} $ are the outer and inner horizon of the R-N black hole. According to the standard ansatz, the wave function of the corrected Dirac equation is always written as\cite{res}
\begin{equation}\label{q5}
\Psi = \varepsilon (t, x^j) \exp{[\frac{i}{\hbar} S(t, x^j)]}.
\end{equation}
Here, both $ \varepsilon $ and $ S $ are the functions of the coordinates $({t,{x^j}})$, and $ S $ is the action of the emission fermion. Substituting the wave function (\ref{q5}) into the corrected Dirac equation (\ref{q2}), and carrying on the separation of variables as $ S = - \omega t + W(r) + \Theta(\theta,\varphi) $ for the spherically symmetric R-N spacetime\cite{ZXX1,ZXX2,ZXX3,ZXX5,ZXX4}, we have
\begin{equation}\label{q6}
\begin{aligned}
\left[ {i{\gamma ^\mu }\left( {{\partial_\mu }S + e{A_\mu }} \right) + m - il_p^{1/2}{\gamma^t}\left( {\omega- e{A_t}} \right){\gamma^j}\left( {{\partial_j}S + e{A_j}} \right)} \right]\\
\times \varepsilon \left( {t,r,\theta,\varphi } \right) = 0,
\end{aligned}
\end{equation}
where, the terms that are related to the high orders of $ \hbar $ are neglected by considering the WKB approximation, $ \hbar \Omega_\mu $ has also been ignored at high energy levels, and $ \omega $ is the energy of the emission fermion. It is well known that, there are two states for the spin-1/2 particles, which correspond, respectively, to the spin up-$ \varepsilon_{\uparrow}(t,r,\theta,\varphi) $ and spin down-$ \varepsilon_{\downarrow}(t,r,\theta,\varphi) $ states. Without loss of generality, it is enough for us to choose the spin-up state. In this case, we have
\begin{equation}\label{q7}
\varepsilon_{\uparrow} \left( {t,r,\theta ,\varphi } \right) = \left( {\begin{array}{*{20}{c}}
{A_{\uparrow}\left( {t,r,\theta ,\varphi }\right)\zeta_{\uparrow}}\\
{B_{\uparrow}\left( {t,r,\theta ,\varphi } \right)\zeta_{\uparrow}}\\
\end{array}} \right).
\end{equation}
where $ \zeta_{\uparrow}=\Big( {\begin{array}{*{20}{c}}
{1}\\
{0}\\
\end{array}} \Big) $ for the spin-up state. To solve the equation (\ref{q6}), the choice of the suitable gamma matrixes is very important. There are many choices to construct the $ \gamma $ matrices, and in this paper we employ
\begin{equation}\label{q8}
\begin{array}{l}
{\gamma ^t} = \frac{1}{{\sqrt f}}\left( {\begin{array}{*{20}{c}}
0&I\\
{ - I}&0
\end{array}} \right),\qquad {\gamma^r} = \sqrt g \left( {\begin{array}{*{20}{c}}
0&{{\sigma^3}}\\
{{\sigma^3}}&0
\end{array}} \right)\\
{\gamma^\theta} = \sqrt {{g^{\theta \theta}}} \left( {\begin{array}{*{20}{c}}
0&{{\sigma^1}}\\
{{\sigma^1}}&0
\end{array}} \right),\qquad {\gamma^\varphi } = \sqrt {{g^{\varphi \varphi}}} \left( {\begin{array}{*{20}{c}}
0&{{\sigma^2}}\\
{{\sigma^2}}&0
\end{array}} \right)
\end{array},
\end{equation}
where, $ {\sigma^i} $ are the Pauli matrixes with $ {i = 1,2,3} $. Inserting the function $ \varepsilon_{\uparrow} \left({t,r,\theta,\varphi} \right) $ and $ \gamma $ matrices into the generalized Dirac equation (\ref{q6}), four simplified equations which are related to the functions $ (A,B) $ can be obtained, and two of them are
\begin{equation}\label{q9}
\begin{aligned}
B\left( { - \frac{{i(\omega-e A_t) }}{{\sqrt f }} + i\sqrt g {\partial _r}W} \right) \qquad \qquad \qquad \qquad \\
 = - A\left( {m - il_p^{1/2}(\omega-e A_t) {\partial _r}W} \right),
\end{aligned}
\end{equation}
\begin{equation}\label{q10}
\begin{aligned}
A\left( {\frac{{i(\omega-e A_t) }}{{\sqrt f }} + i\sqrt g {\partial _r}W} \right) \qquad \qquad \qquad \qquad \\
= - B \left( {m + il_p^{1/2}(\omega-e A_t) {\partial _r}W} \right).
\end{aligned}
\end{equation}
Obviously, the functions $ A $ and $ B $ are required here to keep a non-trivial solution, which demands that the determinant of the coefficient matrix should be zero. In this case, it yields
\begin{equation}\label{q11}
\begin{aligned}
{\partial _r}W( r ) &=  \pm \sqrt {\frac{{\frac{{{(\omega - e A_t) ^2}}}{{{f^2}}} - \frac{{{m^2}}}{f}}}{{( {1 + {{{l_p}{(\omega - e A_t) ^2}} \mathord{/
 {\vphantom {{{l_p}{(\omega - e A_t) ^2}} f}}
 \kern-\nulldelimiterspace} f}} )}}} \\ &=  \pm \sqrt { {\frac{{{(\omega - e A_t) ^2}}}{{{f^2}}} - \frac{{{m^2}}}{f}}} \left( 1 - \frac{l_p(\omega - e A_t)^2}{2 f} \right)
 \end{aligned},
\end{equation}
where the high order terms of $ {l_p} $, ie. $ O\left({\ge l_p^2} \right) $ is so small that it has been neglected. By using the residue principle near the event horizon of the R-N black hole \cite{res}, the value of $ W( r ) $ reads
\begin{equation}\label{q12}
W _{\pm}(r)=  \pm i \pi \left( \frac{r_+^2}{r_+ - r_-} \right)(\omega - e A_{t+})( {1 - {l_p} \Xi}).
\end{equation}
Here, the signs $ \pm $ corresponds to the solutions of the outgoing(ingoing) particles, and $ A_{t+} = \frac{Q}{r_+} $ is the electromagnetic potential at the event horizon, and the parameter $\Xi$ is given by $
\Xi = \frac{e Q \omega (5 r_- - r_+)}{2(r_+ - r_-)^2} +\frac{r_+ \omega^2 (r_+ - 2r_-)}{(r_+ - r_-)^2}  -\frac{m^2}{4}
-\frac{e^2 Q^2}{2 r_+ (r_+ - r_-)}$. Following the WKB approximation, it is well known that the relationship between the imaginary part of the action and the tunneling probability can be expressed as $ P = \exp ({-\frac{2}{\hbar }{\mathop{\rm Im}\nolimits}S})$\cite{Jiang11,Chen1}, so the total emission rate of the Dirac particles can be written as
\begin{equation}\label{q14}
\begin{aligned}
\Gamma &= \frac{{{P_{out}}}}{{{P_{in}}}} = \frac{{\exp ({- 2{\mathop{\rm Im}\nolimits} {W_+ }})}}{{\exp({-2{\mathop{\rm Im}\nolimits} {W_-}})}} \\
&= \exp \left[{-4 \pi \left( \frac{r_+^2}{r_+ - r_-}\right) (\omega -e A_{t+}) ({1-{l_p} \Xi}})\right].
\end{aligned}
\end{equation}
Clearly, there is a small correction to the semiclassical tunneling rate when the effect of the LIV-DR is included. As defined by \cite{ChenT,VagenasT,Jr13T}, after using the principle of ``detailed balance" for the emission rate (\ref{q14}), the effective temperature of the R-N black hole is given by
\begin{equation}\label{q15}
T = \frac{r_+ - r_-}{{4 \pi {r_+}^2 ( {1 - {l_p} \Xi})}} = {T_0}( {1 + {l_p} \Xi }),
\end{equation}
where ${l_p} = \frac{1}{({\xi_2}^2 {M_p}^2)} =\frac{{L_p}^2}{{\xi_2}^2}$, and $ T_0 = \frac{r_+ - r_-}{4 \pi {r_+}^2}$ is the standard Hawking temperature of the R-N black hole and the other terms come from the corrections due to the effect of the LIV-DR. From Eq.(\ref{q15}), we at first sight find the black hole radiation is not only related to black hole parameters $(M,Q)$, but also to the energy $(\omega)$, the charge $(e)$ and the mass $(m)$ of the emission particle. In the GUP case, this similar result has been also produced in \cite{NozariJHEP,ChenT} and references therein.
With the aid of the equation (\ref{q15}), we can plot the Figs.\ref{fig4},\ref{fig1} to visually show the LIV-DR induced corrections to the emission rate and effective temperature of the R-N black hole versus the charge $e$ and energy $\omega$ of the emitted particle.
\begin{center}
\includegraphics[width=6cm]{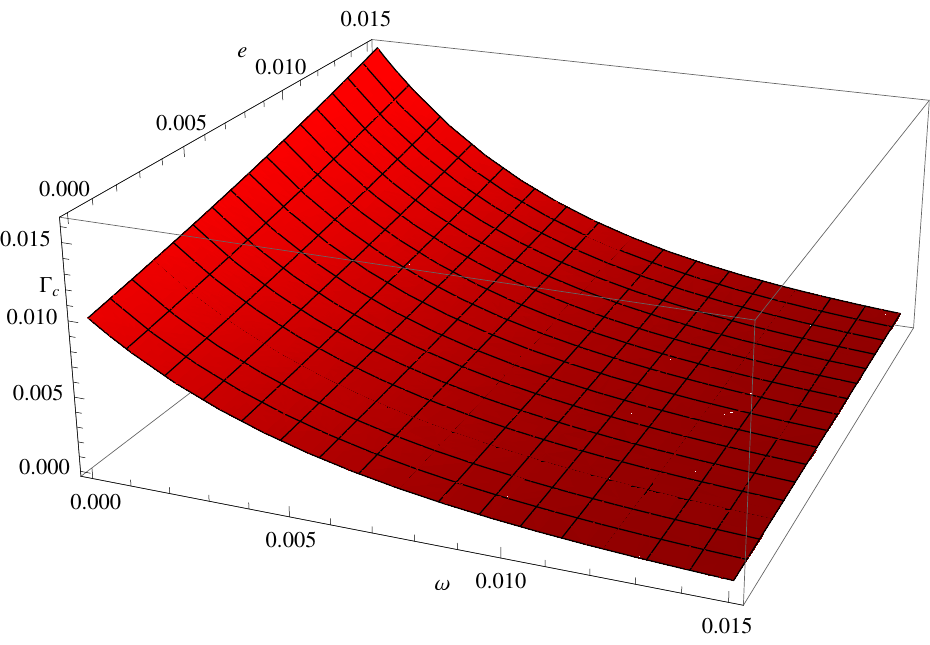}
\figcaption{\label{fig4}  The LIV-DR induced correction to the emission rate of the R-N black hole versus the charge $e$ and energy $\omega$ of the emitted particle, i.e., $\Gamma_c = - l_p \Gamma_0 \Xi$, where $\Gamma_0$ is the original emission rate of the Dirac particles of the R-N black hole. Here we have employed some acceptable parameters, i.e., $ M=30, Q=10, l_p=0.01, m=1,c=1, k_B=1$.}
\end{center}
In Fig.\ref{fig4}, it is easy to see that the LIV induced correction of the emission rate increases with the parameter $e$, but decreases with the particles' energy $\omega$. More importantly, the result also shows that the quantum gravity induced LIV correction of the emission rate is always a positive value, which means that quantum gravity give rise to the increase of the tunneling probability of emitted Dirac particles. So, we can conclude that quantum gravity speeds up the black hole evaporation.
\begin{center}
\includegraphics[width=6cm]{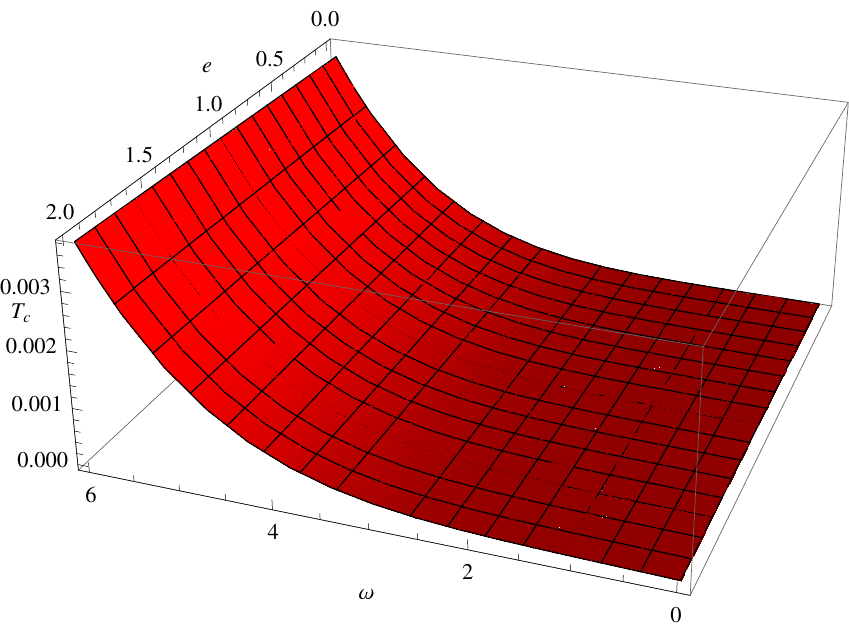}
\figcaption{\label{fig1}  The LIV-DR induced correction to the effective temperature of the R-N black hole versus the charge $e$ and energy $\omega$ of the emitted particle, i.e., $T_c = l_p T_0 \Xi$. Here we have employed some acceptable parameters, i.e., $ M=30, Q=28, l_p=0.01, m=1,c=1, k_B=1$.}
\end{center}
In Fig.\ref{fig1}, it is obvious that the LIV-DR induced correction to the effective temperature of the R-N black hole is bigger and bigger with the increase of the emitted particle's energy $\omega$, but is more or less unchanged with the increase of the emitted particle's charge $e$ at a certain energy level. In addition, gravity as a special interaction force, is very sensitive to the mass of the particle, i.e. the energy of the particle according to the mass-energy relation $\omega=mc^2$. This phenomenon is distinguished from several other interaction forces (i.e. electromagnetic interaction, strong interaction and weak interaction). Combined with theses facts, Fig.\ref{fig1} can give us a phenomenological evidence for the LIV-DR as a candidate for describing the effect of quantum gravity.

To put our results in a proper perspective, we have compared with the earlier findings by the another modified dispersion relation that demands a minimum measurable length and a maximum measurable momentum (i.e. MDR) \cite{KamaliGRG,MajhiPLB}. In Ref. \cite{KamaliGRG}, if we only consider the 4-dimensional R-N black hole, and keep terms up to order of $\alpha$, the corrected temperature reads $T = T_0 \big(1+{2\alpha}/{(M + \sqrt{M^2 - Q^2}})^2\big)$. Here, comparing the effective temperature induced by the MDR with that induced by the LIV-DR, we can plot Fig.\ref{fig2} \footnote{Here, we have replaced $\omega$ in the equation (\ref{q15}) with the characteristic energy of the emitted particle with $T$, i.e. $\omega=k_B T$ \cite{RabinPLB}. }

\begin{center}
\includegraphics[width=6cm]{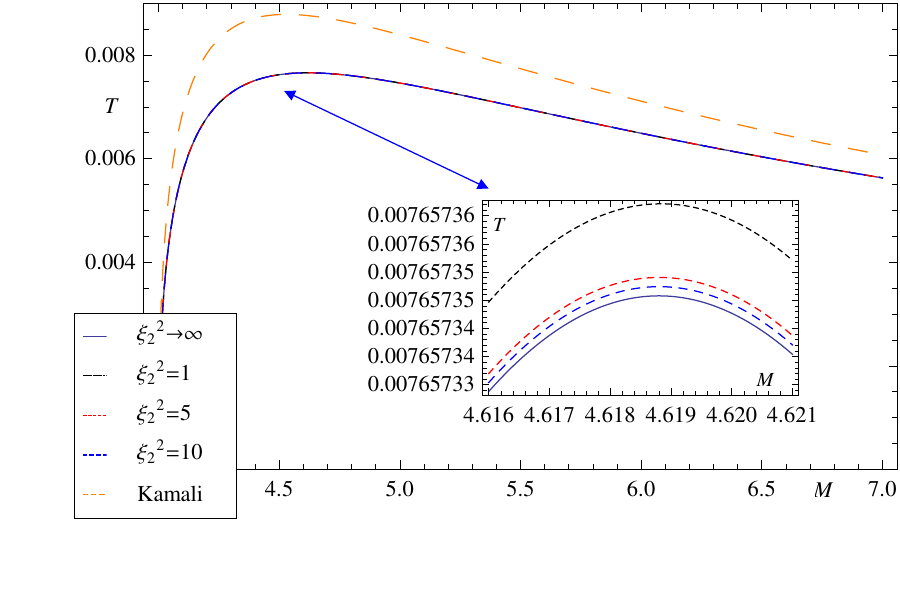}
\figcaption{\label{fig2}   The effective temperature of the R-N black hole versus its mass $M$ for different values of $\xi^2_2$ and $\alpha=1$.  For simplify, we have employed some acceptable parameters, i.e., $Q=4, M_p=1, k_B=1, c=1, m=0.01, e=0.01,\alpha=1 $. }
\end{center}
\begin{center}
\includegraphics[width=6cm]{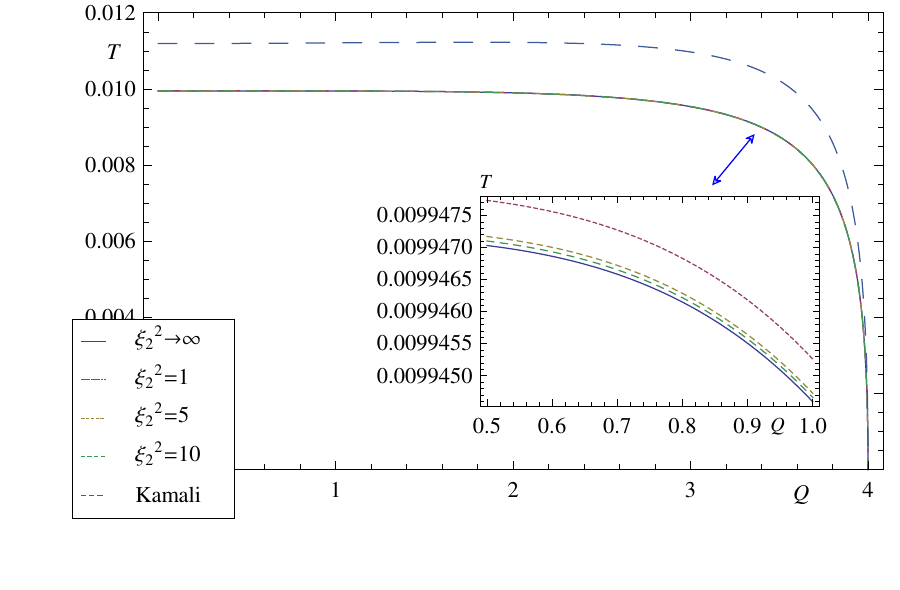}
\figcaption{\label{fig3}   The effective temperature of the R-N black hole versus its charge $Q$ for different value of $\xi^2_2$ and $\alpha=1$. For simplify, the value of $M$ has been set into 4. }
\end{center}
In Figs.\ref{fig2} and \ref{fig3}, we have chosen some reasonable values of $\xi_2$, which are within a range of parameters from flaring active galactic nucleus (AGNs) for $n = 2$ case $\xi_2 \geq 10^{-9}$. In Figs.\ref{fig2} and \ref{fig3}, we first find the case $\xi^2_2 \rightarrow \infty$ corresponds to the standard Hawking temperature of the R-N black hole. Also, we find both the effects of the MDR and the LIV-DR speed up the black hole evaporation respectively. But the MDR-induced departure from the standard temperature is much larger than the LIV-DR induced one when the model-dependent parameter $\xi^2_2=\alpha$. The reason for the tracing is that the dispersion relation of the MDR is linearly suppressed by the power $(E/E_{QG})$, yet that of the LIV-DR is quadratically suppressed by the power $(E/E_{QG})^2$.

In a word, the effective temperature (\ref{q15}), obtained by semi-classical tunneling method, shows Planck-scale correction to original black hole temperature, which is a result of the LIV-DR effect. Meanwhile, we note that the similarly corrected temperature has also been presented in the context of GUP model \cite{NozariJHEP,ChenT,NouicerPLB,NouicerPLB1,NouicerPLB2,Chenpr}. It is found in GUP model that, such a temperature is an important result, which may not only show some intriguing properties at the final stage of the black hole evaporation \cite{ChenT,NouicerPLB,NouicerPLB1,NouicerPLB2,Chenpr}, but also give some suggestions to the so-called black hole information loss paradox \cite{NozariJHEP,Chenpr}. In our previous study\cite{LGP1}, effect of quantum gravity on black hole thermodynamics has been carefully addressed, and some interesting results are obtained. However, in the context of the LIV-DR, whether there exists some intriguing properties about information loss remains unknown. In view of this, it is very interesting for us to further discuss the black hole information loss problems when including the effect of the LIV-DR.

\section{Information loss and entropy conservation}\label{sec3}

In this section, we attempt to investigate the black hole information loss problems within the inclusion of the LIV-DR effect. For simplicity, assuming that the black hole is uncharged (i.e.$Q=0$), then the R-N spacetime naturally reduced to the Schwarzschild spacetime. In this case, as described in our research\cite{LGP1}, it shows that, when the heat capacity of the black hole is equal to zero, the black hole would stop a further collapse at a remnant mass, temperature and entropy, i.e.,
\begin{equation}\label{q18}
\begin{aligned}
{{M}_{rem}} &= \frac{{{M_p}}}{{4\pi \xi_2}},\\
{{T}_{rem}} &= \frac{{\xi _2 {M_p}}}{{{k_B}}}, \\
{{S_{rem}}} &= \frac{k_B}{16 \pi \xi_2^2}\left(1 - \ln \frac{1}{\xi_2^2}\right)
\end{aligned}
\end{equation}
Furthermore, the quantum-gravity induced black hole entropy can also be written in a familiar form,
\begin{eqnarray}\label{s1}
\frac{S}{k_B}&=&\frac{\widetilde{\textbf{A}}}{4L_p^2}-\frac{1}{16\pi \xi_2^{2}}ln \Big(\frac{\widetilde{\textbf{A}}}{4L_p^2}\Big)
+\sum_{j=0}^\infty c_j(\xi_2^{-2})\Big(\frac{\widetilde{\textbf{A}}}{4L_p^2}\Big)^{-j}\nonumber\\
&-&\frac{1}{16\pi \xi_2^{2}} ln 16\pi. \label{Jeq22}
 \end{eqnarray}
where the coefficients $c_j$ are the functions about $\xi_2^{-2}$. The new variable $\widetilde{\textbf{A}}$ is defined by $\widetilde{\textbf{A}}=16\pi G^2 M^2-\frac{2}{4\pi \xi_2^{2} }G^2M_p^2=\textbf{A}-\frac{2}{4\pi \xi_2^{2} } L_p^2$, which is the reduced area, and $\textbf{A}=16\pi G^2 M^2$ is the usual area of the Schwarzschild black hole horizon. Obviously, it is easy to see that the equation (\ref{s1}) can be regarded as a new area theorem by considering the quantum gravity effect, which is similar to the standard modified area theorem\cite{QG1,VagenasT,area1,area2,area3,area4,field}. Most importantly, we find that
LIV-DR produced a Logarithmic correction to the black hole entropy, which is full in consistence with that obtained in \cite{loop,field,quantum,sta}. In a word, the effect of LIV-DR on black hole thermodynamics coincides with that found in another quantum gravity candidate(i.e. GUP model), which implies that the LIV-DR is also a good effective phenomenological model of quantum gravity.
Meanwhile, the GUP model, that gives rise to a nonthermal spectrum to black hole radiation, has shown some statistical correlations with Planck correction between quanta of Hawking radiation \cite{NozariJHEP}. It was found that this correlation can provide a possible solution to the information loss paradox. However, they are not adequate by themselves to recover the information. In \cite{NozariJHEP}, it should be pointed out that the influence of conditional probability and the question that whether the entropy and information are conserved or not were lacking.
Motivated by this fact, it is very interesting for us to detail the information loss problems in the context of the LIV-DR.

The information loss paradox during the Hawking radiation is an outstanding issue for the black hole physics. To date some attempts on this progress have been proposed to solve this paradox \cite{Chenpr,NozariJHEP,ZhangPLB,ZhangPLB1,ZhangPLB2,lf3,lf11,lf4,lf1,infor22,infor33,infor1,infor2,infor3,infor31,infor32,infor333,infor34,infor35,infor36,infor37,
addinfor1,addinfor2}. In particular, using the nonthermal radiation spectrum that originated from the self-gravitational effect, an interesting observation has shown that the Hawking radiation as tunneling is an entropy conservation process, which leads naturally to the conclusion that the process of Hawking radiation is unitary, and no information loss occurs \cite{ZhangPLB,ZhangPLB1,ZhangPLB2}. However, the effect of quantum gravity was lacking in resolving this paradox. To exactly solve the information loss paradox, the quantum gravity effect and the self-gravitational effect should be all considered.
In view of this, before we discuss the information loss problems, it is necessary to reproduce the tunneling probability with inclusion of the self-gravitational effect in the first place. If the self-interaction effect is taken into account \cite{KrausNPB}, the tunneling rate in the presence of the LIV-DR can be obtained with the aid of the relationship $\Gamma \sim \exp[\Delta S] = \exp[S_{(M-\omega)}-S_{(M)}]$ \cite{lf3}, that is
\begin{equation}\label{TR}
\begin{aligned}
\Gamma = &\left( \frac{ \widetilde{\textbf A}_{(M-\omega)}}{\widetilde{\textbf A}_{(M)}}\right)^{\frac{-1}{16 \pi \xi_2^2}} \cdot \exp \left[- 8 \pi \omega \left( M - \frac{\omega}{2}\right)\right] \\
&\cdot \exp \left[   \sum ^{\infty} _{j=0} c_j(\xi_2^{-2}) \left( \frac{{\widetilde{\textbf A}_{(M-\omega)}^{-j}} - {\widetilde{\textbf A}_{(M)}^{-j}}}{4^{-j}} \right) \right]
\end{aligned},
\end{equation}
where, $\widetilde{\textbf A}_{(M-\omega)}=16 \pi (M - \omega)^2 - \frac{2}{4 \pi \xi_2^2}$, and $\widetilde{\textbf A}_{(M)}=16 \pi M^2 - \frac{2}{4 \pi \xi_2^2}$. For convenience, here we have used units $G = k_B = L_p = 1$. The expression (\ref{TR}), obtained by semi-classical tunneling method, shows deviation from the thermal spectrum radiation, which is a result of the self-interaction and the LIV-DR. Using this expression, we will detail the information loss problems in the next step.

For a black hole with the initial mass $M$, if one considers a successive emission with an energy $E_1$, the associated probability can be expressed as \cite{ZhangPLB,ZhangPLB1,ZhangPLB2},
\begin{equation}\label{q22}
\begin{aligned}
\Gamma(E_1) = &\left( \frac{ \widetilde{\textbf A}_{(M-E_1)}}{\widetilde{\textbf A}_{(M)}}\right)^{\frac{-1}{16 \pi \xi_2^2}} \cdot \exp \left[- 8 \pi E_1 \left( M - \frac{E_1}{2}\right)\right] \\
&\cdot \exp \left[   \sum ^{\infty} _{j=0} c_j(\xi_2^{-2}) \left( \frac{{\widetilde{\textbf A}_{(M-E_1)}^{-j}} - {\widetilde{\textbf A}_{(M)}^{-j}}}{4^{-j}} \right) \right]
\end{aligned}.
\end{equation}
For sequential emissions of energies $E_1$ and $E_2$, the tunneling probability for the second emission with an energy $E_2$ should be understood as the conditional probability given the occurrence of tunneling of the particle with an energy $E_1$. In this sense, we have
\begin{equation}\label{q23}
\begin{aligned}
&\Gamma(E_2 \mid E_1) = \left( \frac{ \widetilde{\textbf A}_{(M-E_1-E_2)}}{\widetilde{\textbf A}_{(M-E_1)}}\right)^{\frac{-1}{16 \pi \xi_2^2}} \\
&\cdot \exp \left[- 8 \pi E_2 \left( M - E_1 - \frac{E_2}{2}\right)\right] \\
&\cdot \exp \left[   \sum ^{\infty} _{j=0} c_j(\xi_2^{-2}) \left( \frac{{\widetilde{\textbf A}_{(M-E_1-E_2)}^{-j}} - {\widetilde{\textbf A}_{(M-E_1)}^{-j}}}{4^{-j}} \right) \right]
\end{aligned}.
\end{equation}
The probability for simultaneously two emissions with energies $E_1$ and $E_2$ is
\begin{equation}\label{q24}
\begin{aligned}
&\Gamma(E_1+E_2) = \left( \frac{ \widetilde{\textbf A}_{(M - E_1 - E_2)}}{\widetilde{\textbf A}_{(M)}}\right)^{\frac{-1}{16 \pi \xi_2^2}} \\
&\cdot \exp \left[- 8 \pi ( E_1 + E_2) \left( M - \frac{E_1+E_2}{2}\right)\right] \\
&\cdot \exp \left[   \sum ^{\infty} _{j=0} c_j(\xi_2^{-2}) \left( \frac{{\widetilde{\textbf A}_{(M-E_1-E_2)}^{-j}} - {\widetilde{\textbf A}_{(M)}^{-j}}}{4^{-j}} \right) \right]
\end{aligned}.
\end{equation}
Using the standard approach that described in \cite{ZhangPLB,ZhangPLB1,ZhangPLB2}, the independent probability for the second emission is taken the expected functional form of equation (\ref{TR}), that is,
\begin{equation}\label{second}
\begin{aligned}
\Gamma(E_2 ) &= \left( \frac{ \widetilde{\textbf A}_{(M-E_2)}}{\widetilde{\textbf A}_{(M)}}\right)^{\frac{-1}{16 \pi \xi_2^2}} \cdot \exp \left[- 8 \pi E_2 \left( M - \frac{E_2}{2}\right)\right] \\
&\cdot \exp \left[   \sum ^{\infty} _{j=0} c_j(\xi_2^{-2}) \left( \frac{{\widetilde{\textbf A}_{(M-E_2)}^{-j}} - {\widetilde{\textbf A}_{(M)}^{-j}}}{4^{-j}} \right) \right]
\end{aligned}.
\end{equation}
For two emissions ($E_1, E_2$), it is obvious that their joint probability $\Gamma(E_1, E_2)$\footnote{Where, $\Gamma(E_1, E_2) = \Gamma(E_1+ E_2).$} is not equal to the sum of the each single emissions ($\Gamma(E_1), \Gamma(E_2)$) of Hawking radiation, i.e., $\Gamma(E_1+ E_2) \neq \Gamma(E_1) + \Gamma(E_2)$. Alternatively, one can find that the relationship $\Gamma(E_1,E_2) = \Gamma(E_1) \cdot \Gamma(E_2 \mid E_1) = \Gamma(E_1 + E_2) $ obtained in \cite{ZhangPLB,ZhangPLB1,ZhangPLB2} without including the effect of quantum gravity also holds true here. In view of this, it is true that the nontrivial correlation must exist between two sequential emissions ($E_1, E_2$), and they are indeed dependent. To present this correlation, a quantity that used to measure the correlation between sequential emissions $E_1$ and $E_2$ has been defined in Ref.\cite{ZhangPLB,ZhangPLB1,ZhangPLB2}, which is
\begin{equation}\label{Cd}
\begin{aligned}
&\mathcal{C}(E_1+E_2;E_1,E_2)\\
&= \ln \Gamma(E_1+E_2) - \ln \Gamma(E_1) - \ln \Gamma(E_2)
\end{aligned}.
\end{equation}
With aid of Eqs. (\ref{q22}),(\ref{q24}),(\ref{second}), the correspondingly correlation function $\mathcal{C}(E_1+E_2;E_1,E_2)$ between the two emitted particles is easy to get, which is
\begin{equation}\label{q26}
\begin{aligned}
&\mathcal{C}(E_1+E_2;E_1,E_2) \\
&= 8\pi {E_1}{E_2} - \frac{1}{16\pi \xi_2^2} \ln \left( \frac{\widetilde{\textbf A}_{(M-E1-E_2)} \widetilde{\textbf A}_{(M)}}{\widetilde{\textbf A}_{(M-E1)} \widetilde{\textbf A}_{(M-E_2)}} \right)\\
& + \sum ^{\infty} _{j=0} c_j(\xi_2^{-2}) \frac{  {\widetilde{\textbf A}_{(M-E_1-E_2)}^{-j}} + {\widetilde{\textbf A}_{(M)}^{-j}}   }{4^{-j}}\\
& - \sum ^{\infty} _{j=0} c_j(\xi_2^{-2}) \frac{  {\widetilde{\textbf A}_{(M-E1)}^{-j}} + {\widetilde{\textbf A}_{(M-E2)}^{-j}}   }{4^{-j}}
\end{aligned}.
\end{equation}
By comparing with the correlation functions obtained in \cite{ZhangPLB,ZhangPLB1,ZhangPLB2}, we found that there is a new term with quantum gravity correction, enhances the statistical correlations in our case due to the corrected nonthermal spectra produced by the LIV-DR. This result is compatible with the finding that obtained in Ref.\cite{NozariJHEP}. Meanwhile, we can continue to calculate the correlations between those $n-1$ emissions and the $n$th emission with energy $E_n$. Therefore, by considering the remnant value of black hole is $M_{rem}$ in unit $c=1$ \footnote{Where $ M $ is the initial mass of black hole.}, then the total correlation among a queue of Hawking radiations $E_1,E_2,\cdots E_n$ can be summed up to
\begin{equation}\label{coradd}
\begin{aligned}
&\mathcal{C}(M - M_{rem};E_1,E_2 \cdots E_n) \\
&= \sum_{n \geqslant 2} 8\pi ({E_1 + E_2 + \cdots + E_{n-1}}){E_n} \\
&- \frac{1}{16\pi \xi_2^2} \ln \left( \frac{\widetilde{\textbf A}_{(M-M_{rem})} \widetilde{\textbf A}_{(M)}^{n-1} }{\widetilde{\textbf A}_{(M-E1)} \widetilde{\textbf A}_{(M-E_2)}\cdots \widetilde{\textbf A}_{(M-E_n)}} \right)\\
& + \sum ^{\infty} _{j=0} c_j(\xi_2^{-2}) \frac{  {\widetilde{\textbf A}_{(M-M_{rem})}^{-j}} + (n-1){\widetilde{\textbf A}_{(M)}^{-j}}   }{4^{-j}}\\
& - \sum ^{\infty} _{j=0} c_j(\xi_2^{-2}) \frac{  {\widetilde{\textbf A}_{(M-E1)}^{-j}} + {\widetilde{\textbf A}_{(M-E2)}^{-j}} + \cdots {\widetilde{\textbf A}_{(M-E_n)}^{-j}}  }{4^{-j}}
\end{aligned},
\end{equation}
where, the relation $M = \sum\limits^{i=n} _{i=1} E_i + M_{rem}$ has been used in above calculation. From Eq.(\ref{coradd}), one can see that when the parameter $\xi_2 \rightarrow \infty$ the correlation will naturally reduced to $\sum\limits_{n \geqslant 2} 8\pi ({E_1 + E_2 + \cdots + E_{n-1}}){E_n}$, which is full in consistence with that gotten by B.C. Zhang \cite{addinfor1,addinfor2}. On the other hand, the concept of the mutual information in a composite quantum system composed of sub-systems $A$ and $B$ is defined as $S(A:B) = S(A) + S(B) - S(A,B) = S(A) - S(A \mid B)$, where $S(A \mid B)$ is the conditional entropy. As described by \cite{addinfor1,addinfor2}, this information $S(A:B)$ can be used to measure the total correlations between the any bi-partite systems. In this sense, with inclusion of the effect of quantum gravity, we find that the mutual information for sequential emission of two emissions ($E_1$ and $E_2$) is exactly equal to the correlation of Eq.(\ref{q26}), which makes a reasonable understanding of the correlation (\ref{q26}). However, this nontrivial correlation (\ref{coradd}) is far away from enough if one attempts to recover the black hole information. So, it is necessary for us to carefully reexamine the entropy and information conservation in the next step.
\\
\indent
For the first particle with an energy $E_1$, when it emitted from a black hole with a mass $ M $, the entropy that carried away by the emission $E_1$ is
\begin{equation}\label{q27}
\begin{aligned}
S(E_1) = - \ln \Gamma( E_1 )
\end{aligned}.
\end{equation}
As sequential emissions, the conditional entropy that carried away by the second emission $E_2$ is
\begin{equation}\label{q28}
\begin{aligned}
S(E_2 \mid E_1)= - \ln \Gamma( E_2\mid E_1 )
\end{aligned}.
\end{equation}
Therefore, the total entropy that carried away by the two emitted particles can be expressed as
\begin{equation}
 S(E_1,E_2) = S(E_1) + S(E_2\mid E_1).
\end{equation}
Repeating the process until the black hole radiation ceased, then we can easily find the total entropy that carried away by all emissions is \cite{ZhangPLB,ZhangPLB1,ZhangPLB2}
\begin{equation}\label{q29}
\begin{split}
S(E_1,E_2,\cdots,E_n) &= \sum \limits ^n_{i=1}S(E_i \mid E_1,E_2,\cdots,E_{i-1})\\
 &= - \ln \prod_{i=1}^{n} \Gamma \left( M - \sum \limits ^{i-1}_{j=1} E_j ; E_i \right)
\end{split}.
\end{equation}
with
\begin{equation}\label{q30}
\begin{aligned}
&\Gamma(M;E_1) = \left( \frac{ \widetilde{\textbf A}_{(M-E_1)}}{\widetilde{\textbf A}_{(M)}}\right)^{\frac{-1}{16 \pi \xi_2^2}} \cdot \exp \left[- 8 \pi E_1 \left( M - \frac{E_1}{2}\right)\right] \\
&\cdot \exp \left[   \sum ^{\infty} _{j=0} c_j(\xi_2^{-2}) \left( \frac{{\widetilde{\textbf A}_{(M-E_1)}^{-j}} - {\widetilde{\textbf A}_{(M)}^{-j}}}{4^{-j}} \right) \right],\\
&\Gamma(M-E_1;E_2) = \left( \frac{ \widetilde{\textbf A}_{(M-E_1-E_2)}}{\widetilde{\textbf A}_{(M-E_1)}}\right)^{\frac{-1}{16 \pi \xi_2^2}} \\
&\cdot \exp \left[- 8 \pi E_2 \left( M - E_1 - \frac{E_2}{2}\right)\right] \\
&\cdot \exp \left[   \sum ^{\infty} _{j=0} c_j(\xi_2^{-2}) \left( \frac{{\widetilde{\textbf A}_{(M-E_1-E_2)}^{-j}} - {\widetilde{\textbf A}_{(M-E_1)}^{-j}}}{4^{-j}} \right) \right],\\
&\cdots \cdots \cdots \cdots ,\\
&\Gamma( M-{\sum \limits ^{n-1}_{i=1} E_i};En) = \left( \frac{ \widetilde{\textbf A}_{(M-{\sum \limits ^{n}_{i=1} E_i})}}{\widetilde{\textbf A}_{(M-{\sum \limits ^{n-1}_{i=1} E_i})}}\right)^{\frac{-1}{16 \pi \xi_2^2}} \\
&\cdot \exp \left[- 8 \pi E_n \left( M - {\sum \limits ^{n-1}_{i=1} E_i} - \frac{E_n}{2}\right)\right] \\
&\cdot \exp \left[   \sum ^{\infty} _{j=0} c_j(\xi_2^{-2}) \left( \frac{{\widetilde{\textbf A}_{(M-{\sum \limits ^{n}_{i=1} E_i})}^{-j}} - {\widetilde{\textbf A}_{(M-{\sum \limits ^{n-1}_{i=1} E_i})}^{-j}}}{4^{-j}} \right) \right],
\end{aligned}
\end{equation}
where $\Gamma(M-E_1;E_2)$ means the probability equation (\ref{TR}) for a emission with energy $E_2$ by a black hole with mass $(M-E_1)$, which is same as the meaning of $\Gamma(E_2|E_1)$. In previous section, we have confirmed that when including the effect of quantum gravity during the emission process, a black hole stops a further collapse at a remnant mass, and becomes an inert remnant. In this sense, black hole remnants should possess a certain amount of entropy at the final stage of black hole evaporation. Assuming that the remnant value of black hole is $M_{rem}$, and considering the relation $M =\sum\limits ^{n}_{i=1} E_i + M_{rem}$. Therefore, the total entropy carried away by all emissions, on the condition that black hole remnant is $M_{rem}$, takes the form,
\begin{equation}\label{q31}
\begin{aligned}
S_{emi} &= 4 \pi (M^2 - M_{rem}^2) - \frac{1}{ 16 \pi \xi_2^2} \ln \frac{\widetilde{\textbf A}_{(M)}}{\widetilde{\textbf A}_{(M_{rem})}} \\
& + \sum ^{\infty} _{j=0} c_j(\xi_2^{-2})   \frac{{\widetilde{\textbf A}_{(M)}^{-j}} - {\widetilde{\textbf A}_{(M_{rem})}^{-j}}}{4^{-j}} \\
&= S_{ini} - S_{rem} = \Delta S
\end{aligned}.
\end{equation}
Where, $S_{ini}$ is the total entropy of an initial black hole, which form is $S_{ini} =  \frac{\widetilde{\textbf A}_{(M)}}{4}  - \frac{1}{ 16 \pi \xi_2^2} \ln \frac{\widetilde{\textbf A}_{(M)}}{4} + \sum_{j=0}^\infty c_j(\xi_2^{-2})\Big(\frac{ {\widetilde{\textbf A}_{(M)}} }{4}\Big)^{-j} - \frac{1}{16 \pi \xi_2^2} \ln 16 \pi$, $S_{rem}$ is the entropy of black hole remnants, it is $S_{rem} =  \frac{\widetilde{\textbf A}_{(M_{rem})}}{4}  - \frac{1}{ 16 \pi \xi_2^2} \ln \frac{\widetilde{\textbf A}_{(M_{rem})}}{4} + \sum_{j=0}^\infty c_j(\xi_2^{-2})\Big(\frac{ {\widetilde{\textbf A}_{(M_{rem})}} }{4}\Big)^{-j} - \frac{1}{16 \pi \xi_2^2} \ln 16 \pi$, and $S_{emi}$ is the entropy of all emitted particles. From equation (\ref{q31}), it turns out that the total entropy ($S_{ini}$) of an initial black hole is equal to the sum of the entropy ($S_{emi}$) carried away by all emissions and the residual entropy $(S_{rem})$ of black hole remnants. So, we conclude that the black hole radiation as tunneling is an entropy conservation process, even if there exists the effect of quantum gravity. This conclusion is compatible with the findings that obtained by \cite{lf3,lf11}.
\\
\indent
However, we note in above discussions that the specific meaning of the entropy carried away by an emission remains unclear, thus it is necessary to continue to investigate this puzzling point with inclusion of quantum gravity induced LIV effects. In Ref.\cite{addinfor1,addinfor2}, B.C. Zhang et al have pointed out that the entropy should be regarded as the uncertainty about the information of the precollapsed configurations of a black hole's forming matter, self-collapsed configurations and the inter-collapsed configurations. Specifically, when the back reaction of emission is taken into account and the effect of quantum gravity is out of consideration, the entropy carried away by a particle $S(E) = 8 \pi E ( M- E/2)$ should be reexpressed as $S(E) = 8 \pi E ( M- E) + (4 \pi E^2 - S_0 )+ S_0$. And, $S_0$, $4 \pi E^2 - S_0 $ and $8 \pi E ( M- E)$ are the inherent entropy of the radiating particle which referring to the precollapsed configuration, the entropy of the remaining black hole which referring to the self-collapsed configuration and the correlation between the radiation and the remaining black hole which referring to the inter-collapsed configuration, respectively. According to this approach, we find in the context of quantum gravity that the entropy carried away by an emission $E$ should be rewritten as
\begin{equation}\label{SS1}
\begin{aligned}
S(E) =\mathcal{S}_{ic} + \mathcal{S}_{sc} + \mathcal{S}_{pc}
\end{aligned},
\end{equation}
with

\begin{equation}\nonumber
\begin{aligned}
\mathcal{S}_{ic} &=  \frac{\widetilde{\textbf A}_M - \widetilde{\textbf A}_{M-E} -\widetilde{\textbf A}_{E}}{4} - \frac{1}{16 \pi \xi_2^2} \ln \frac{4 \widetilde{\textbf A}_M}{\widetilde{\textbf A}_{M-E} \widetilde{\textbf A}_{E}}\\
& + \sum ^{\infty}_{j=0} c_j(\xi_2^{-2}) \frac{\widetilde{\textbf A}^{-j}_M - \widetilde{\textbf A}^{-j}_{M-E} - \widetilde{\textbf A}^{-j}_{E}}{4^{-j}} + \frac{1}{16 \pi \xi_2^2} \ln 16 \pi
\end{aligned},
\end{equation}
\begin{equation}\nonumber
\begin{aligned}
\mathcal{S}_{sc} &=  \frac{\widetilde{\textbf A}_{E}}{4} - \frac{1}{16 \pi \xi_2^2} \ln \frac{\widetilde{\textbf A}_{E}}{4}+ \sum ^{\infty}_{j=0} c_j(\xi_2^{-2}) \frac{\widetilde{\textbf A}^{-j}_{E}}{4^{-j}} \\
&- \frac{1}{16 \pi \xi_2^2} \ln 16 \pi - \mathcal{S}_{pc}
\end{aligned},
\end{equation}
\begin{equation}\label{SS11}
\begin{aligned}
\mathcal{S}_{pc} &= {S}_{0}
\end{aligned}.
\end{equation}
One can easily check from the expression (\ref{SS11}) that the original result i.e., $\mathcal{S}_{ic} = 8 \pi E (M - E), \mathcal{S}_{sc} = 4 \pi E^2 - S_0, \mathcal{S}_{pc} = S_0 $ can be well recovered by setting $\xi_2 \rightarrow \infty$.
For a closed physical system, the information theory tells us that the uncertainty of the event (an emission with an energy $E$) or the information we can gain from this event\cite{final}, on average, is $I(E) = S(E) = -\ln \Gamma(E)$. In this paper, we have emphasized that black hole and its radiations can constitute a closed physics system.
In view of this, the above entropies $\mathcal{S}_{ic}, \mathcal{S}_{sc}$ and $\mathcal{S}_{pc}$ carried away by Hawking radiations
should be interpreted as the uncertainty of information of its inter-collapsed configuration, self-collapsed configuration and precollapsed configuration, respectively.
Meanwhile, we can also employ the the expression (\ref{SS1}) to calculate the total entropy carried away by the sequential Hawking radiation. For the first emitted particle with energy $E_1$, the entropy reads ${S}(E_1) = \mathcal{S}_{ic}(E_1:M-E_1) + \mathcal{S}_{sc}(E_1) + {S}_{01} $. After the first emission, the entropy of the second emitted particle with energy $E_2$ is ${S}(E_2\mid E_1) = \mathcal{S}_{ic}(E_2:M-E_1-E_2) + \mathcal{S}_{sc}(E_2) + {S}_{02}$. Repeating this process until the black hole ceased with the remnant mass $M_{rem}$, the total entropy carried away by black hole radiation can be gotten to be the same as (\ref{q31}), i.e., $S_{emi} = S_{ini} - S_{rem} = \Delta S$. Based on the information theory, the total amount of information carried away by all emissions is $I_{emi}(E_1,E_2,\cdots,E_n) = S( M ) - S\left( M - \sum^{n}_{i=1}E_i\right) = S_{ini} - S_{rem}$. That is to say, the change of entropy denotes the loss of the information of black hole $(\Delta I = - \Delta S)$.
Therefore, it turns out that the entropy and information conservation are preserved at all times. This means all entropy or information of Hawking radiations can be carried away by themselves, and none of which losses. In addition, it is true that when the effect of LIV is included our above discussions can also provide a self-consistent interpretation for the black hole entropy, which is full inconsistence with that obtained in \cite{addinfor1,addinfor2}. Finally, we can conclude in the context of quantum gravity that, the black hole radiation is not only an the entropy and information conservation process, but still an unitary process, which is full in consistent with the findings given by \cite{lf4,lf3,lf11}.

\section{Conclusions and Discussion}\label{sec4}

In this paper, we have applied the Lorentz-invariance-violation (LIV) class of dispersion relations (DR) suppressed by the second power $(E/E_{QG})^2$, to carefully study Hawking radiation of Dirac particles via tunneling from a charged R-N black hole. It turns out that, black hole radiation is not only related to black hole parameters $(M,Q)$, but also to the energy $(\omega)$, the charge $(e)$ and the mass $(m)$ of the emission particle. In Fig.\ref{fig1}, one can see that the LIV-DR induced correction to the effective temperature is bigger and bigger with the increase of the particle's energy $(\omega)$, but has no evidently changes with the emitted particles' charge $(e)$ at certain energy level. This fact provides a phenomenological evidence for the LIV-DR as a candidate for describing the effect of quantum gravity, which once again confirm the conclusion of \cite{Jacob,Jacobson}. Also, to put our results in a proper perspective, we have compared with the earlier findings by another deformed dispersion relations suppressed by the first power $(E/E_{QG})$. In Figs.\ref{fig2},\ref{fig3}, the results show that, the effects of the MDR and the LIV-DR speed up the black hole evaporation respectively. But the MDR-induced departure from the standard temperature is much larger than the LIV-DR induced one when the model-dependent parameter $\xi^2_2=\alpha$.
\\
\indent
The standard Hawking formula predicts the complete evaporation of black holes. However, when including of the effect of quantum gravity during the emission process, we find that there is no complete evaporation at the final stage of black hole evaporation, instead the remnant value of mass ${M_{rem}}$, temperature ${T_{rem}}$ and entropy $S_{rem}$ arise naturally.
In this case, with inclusion of the LIV-DR, we further discuss the black hole information loss problem by using standard statistical method.
The results show that, the original statistical correlation ($\mathcal{C} = 8\pi E_1 E_2$) is modified, and the quantum-gravity-induced LIV-DR produces a new term with Planck-scale correction in statistical correlation function. This effect can not be neglected once the black hole mass becomes comparable to the Planck mass.
Then, based on such correlations, the conditional entropy and total entropy carried away by emitted particles are calculated, where the existence of the residual mass is emphasized at final stage of black hole evolution. Also, we from equation (\ref{q31}) find that the black hole radiation as tunneling is also an entropy conservation process.
Finally, by interpreting the entropy as the uncertainty about information, it shows that Hawking radiation indeed can take away the information and no information loss occurs during the evaporation process.
So, we assert that the black hole evaporation is still an unitary process, even if there exists the effect of quantum gravity.

\setlength{\parindent}{0pt}\textbf{\textbf{Acknowledgments}}
The authors would like to thank Professor Shu-Zheng Yang and Qing-Quan Jiang for their useful discussions, and the anonymous reviewers for their helpful comments and suggestions, which helped to improve the quality of this paper. 
This work is supported by the National Natural Science Foundation of China (Grant No.11903025), and by the starting fund of China West Normal University (Grant No.18Q062), and by the Natural Science Foundation of Sichuan Education Department (Grant No.17ZA0294), and by the Research Project of Si Chuan MinZu College (Grant No.XYZB18003ZA).\\

\end{multicols}
\end{CJK*}
\end{document}